# Thermal Percolation Threshold and Thermal Properties of Composites with Graphene and Boron Nitride Fillers

**Fariborz Kargar[×], Zahra Barani[×], Jacob S. Lewis, Bishwajit Debnath, Ruben Salgado, Ece Aytan, Roger Lake and Alexander A. Balandin[*]**

Phonon Optimized Engineered Materials (POEM) Center, Department of Electrical and Computer Engineering, Materials Science and Engineering Program, Bourns College of Engineering, University of California, Riverside, California 92521 USA

## Abstract

We investigated thermal properties of the epoxy-based composites with a high loading fraction – up to $f \approx 45 \, \text{vol.\%}$ – of the randomly oriented electrically conductive graphene fillers and electrically insulating boron nitride fillers. It was found that both types of the composites revealed a distinctive *thermal* percolation threshold at the loading fraction $f_T > 20 \, \text{vol.\%}$. The graphene loading required for achieving the thermal percolation, $f_T$, was substantially higher than the loading, $f_E$, for the *electrical* percolation. Graphene fillers outperformed boron nitride fillers in the thermal conductivity enhancement. It was established that thermal transport in composites with the high filler loading, $f \geq f_T$, is dominated by heat conduction via the network of percolating fillers. Unexpectedly, we determined that the thermal transport properties of the high loading composites were influenced strongly by the cross-plane thermal conductivity of the quasi-two-dimensional fillers. The obtained results shed light on the debated mechanism of the thermal

---

[×] Contributed equally to the work.
[*] Corresponding author (A.A.B.): balandin@ece.ucr.edu ; web-site: http://balandingroup.ucr.edu/

percolation, and facilitate the development of the next generation of the efficient thermal interface materials for electronic applications.

## Main Text

The discovery of unique heat conduction properties of graphene[1–7] motivated numerous practically oriented studies of the use of graphene and few-layer graphene (FLG) in various composites, thermal interface materials and coatings[8–15]. The intrinsic thermal conductivity of large graphene layers exceeds that of the high-quality bulk graphite, which by itself is very high – $2000\,\mathrm{Wm^{-1}K^{-1}}$ at room temperature (RT)[1,11,16,17]. The first studies of graphene composites found that even a small loading fractions of randomly oriented graphene fillers – up to $f = 10\,\mathrm{vol.\,\%}$ – can increase the thermal conductivity of epoxy composites by up to a factor of ×25 [Ref. 11]. These results have been independently confirmed by different research groups[18,19]. The variations in the reported thermal data for graphene composites can be explained by the differences in the methods of preparation, matrix materials, quality of graphene, lateral sizes and thickness of graphene fillers and other factors[3,20–25]. Most of the studies of thermal composites with graphene were limited to the relatively low loading fractions, $f \leq 10\,\mathrm{vol.\,\%}$. The latter was due to difficulties in preparation of high-loading fraction composites with a uniform dispersion of graphene flakes. The changes in viscosity and graphene flake agglomeration complicated synthesis of the consistent set of samples with the loading substantially above $f = 10\,\mathrm{vol.\,\%}$.

Investigation of thermal properties of composites with the high loading fraction of graphene or FLG fillers is interesting from both fundamental science and practical applications points of view. The high loading is required for understanding the thermal percolation in composites with graphene and other two-dimensional (2D) materials. Thermal percolation in composites, in general, remains a rather controversial issue[26–33]. The electrical percolation in composites with various carbon fillers, including carbon nanotubes and graphene, can be clearly observed experimentally as an abrupt change in the electrical conductivity[34–40]. It is commonly described theoretically by the power scaling law[34–40] $\sigma \sim (f - f_E)^t$, where $\sigma$ is the electrical conductivity of the composite, $f$ is the filler loading volume fraction, $f_E$ is the filler loading fraction at the electrical percolation threshold, and $t$ is the "universal" critical exponent in which $t \approx 2$ represents the percolation in three dimensions[34,36]. However, in most of cases, the thermal conductivity of composites does not reveal such obvious changes as the loading fraction continues to increase. There have been suggestions that thermal percolation threshold does not exist at all[27]. The common argument is that the electrically insulating matrix materials do not conduct electrical current but still conduct heat. Indeed, the ratio of the intrinsic thermal conductivity of graphene fillers, $K_f$, to that of the epoxy matrix, $K_m$, is $K_f/K_m \sim 10^5$. The ratio of the electrical conductivity of the graphene fillers to that of the matrix can be as high as $\sigma_f/\sigma_m \sim 10^{15}$. With such a difference in the ratios, the electrical conduction is only expected via the percolation network while the thermal conduction can still happen through the matrix[27,41]. Even if the thermal percolation is achieved it is still an open question of how much of the heat flux is propagating via the percolated network of fillers in comparison with the thermal transport though the matrix.

There is also a very strong practical motivation for research of composites with the high loading of graphene. There is an increasing demand for better thermal interface materials (TIMs) for heat removal in electronics and energy conversion applications [2,10,42,43]. Commercially available TIMs with the "bulk" thermal conductivity in the range from $\sim 0.5\ \mathrm{Wm^{-1}K^{-1}}$ to $5\ \mathrm{Wm^{-1}K^{-1}}$ no longer meet the industry requirements. Composites with the high loading of graphene fillers have the potential to deliver high thermal conductivity and low thermal contact resistance. Moreover, recent technological developments have demonstrated that the liquid phase exfoliated (LPE) graphene can be produced inexpensively and in large quantities[44,45]. Various methods of reduction of graphene oxide (GO) have also been reported[40,46–48]. The progress in graphene and GO synthesis made FLG fillers practical even for the composites with the high loading fraction. One should note here that FLG fillers with thickness in the range of a few nanometers are substantially different from milled graphite fillers with hundreds of nanometers or micrometer thicknesses. Much thicker graphite fillers do not have the flexibility of FLG and, as a result, do not couple well to the matrix.

In this Letter, we report the results of our investigation of heat transport in the epoxy composites with the high loading fraction – up to $f = 45\,\text{vol.\,\%}$ – of graphene and hexagonal boron nitride ($h$-BN) fillers. The second type of fillers – electrically insulating $h$-BN – was used for comparison with the electrically conducting graphene in order to establish the most *general* trends in thermal conductivity of composites with quasi-2D fillers. We established that both types of the composites revealed a distinctive *thermal* percolation threshold at the loading fraction $f_T \approx 30\,\text{vol.\,\%}$ for graphene and $f_T \approx 23\,\text{vol.\,\%}$ for $h$-BN. The onset of the thermal percolation was achieved at higher loading than the electrical percolation in graphene composites. It was found that the thermal conductivity of both composites in the entire range of loading fractions can be best described by the Lewis-Nielsen model[49,50]. We discovered that contrary to conventional belief the thermal transport in composites with the filler loading $f \geq f_T$ is influenced strongly by the apparent *cross-plane* thermal conductivity of the quasi-2D fillers such as graphene or boron nitride. At low loading, $f \leq f_T$, most of the fillers are not attached to each other, and the thermal transport is governed by the thermal conductivity of the base polymer and the in-plane thermal conductivity of the fillers.

In order to achieve the high loading fraction of fillers, we had to use commercially produced graphene and $h$-BN fillers. In the thermal context, the term "graphene fillers" implies a mixture of graphene and FLG flakes with the lateral size and thicknesses in a certain range, differentiating such fillers from much thicker graphite flakes or nano-platelet graphite powder. The same terminology convention applies to $h$-BN fillers. The composite samples were prepared from the commercially produced graphene flakes (Graphene Supermarket) and $h$-BN flakes (US Research Nanomaterials, Inc.). The lateral size of the few-atomic-layer fillers of graphene ranged from $\sim$2 μm to $\sim$8 μm while the thickness varied from single atomic planes of 0.35 nm to $\sim$12 nm. The lateral dimensions of $h$-BN were comparable, in the interval from $\sim$3 μm to $\sim$8 μm. To obtain a uniform dispersion and avoid air gaps in the high loading composites we used an in-house designed mixer. The graphene and $h$-BN mixtures were not optimized for achieving the largest thermal conductivity enhancement[10,11]. However, the consistent composition used for samples with all loading fractions, allowed us to reveal the percolation trends in such composites.

The composites were prepared in the form of disks with the radius of 25.6 mm and thickness of 3 mm (see Figure 1 (a)). We paid particular attention to accurate determination of the mass density of the resulting samples (Supplementary Figures 1 and 2). The procedures of the sample preparation and characterization are described in details in the Methods section and Supplementary Materials. Below the percolation threshold ($f < f_T$), the fillers do not attach to each other while above it ($f > f_T$) they create a network of pathways for conducting heat (see the schematics in Figure 1 (b)). The samples with different loading fraction of the fillers, $f$, were characterized by the scanning electron microscopy (SEM). Figure 1 (c) shows SEM image of the sample with the high graphene loading ($f \sim 45$ vol. %). One can see clearly the overlapping segments of graphene fillers, indicative of the forming the percolation network. The composition and quality of the graphene and $h$-BN epoxy composites have been verified with Raman spectroscopy (Renishaw InVia). The representative spectra are shown in Figure 1 (d-e). The measurements were performed in the backscattering configuration under visible red laser ($\lambda = 633$ nm). The excitation power on the surface was kept at $\sim 3.6$ mW in all the experiments. In both types of composites, the higher loading of fillers resulted in the increased intensity of the characteristic graphene (G-peak at $\sim 1580$ cm$^{-1}$) or $h$-BN ($E_{2g}$ mode at $\sim 1366$ cm$^{-1}$) phonon peaks, allowing for additional verification of the composition of the samples. It should be noted that in case of epoxy with graphene fillers, the intensity of the 2D peak is much lower than the intensity of the G-peak, indicating the random mixture of single- and few-layer graphene flakes inside the epoxy matrix.

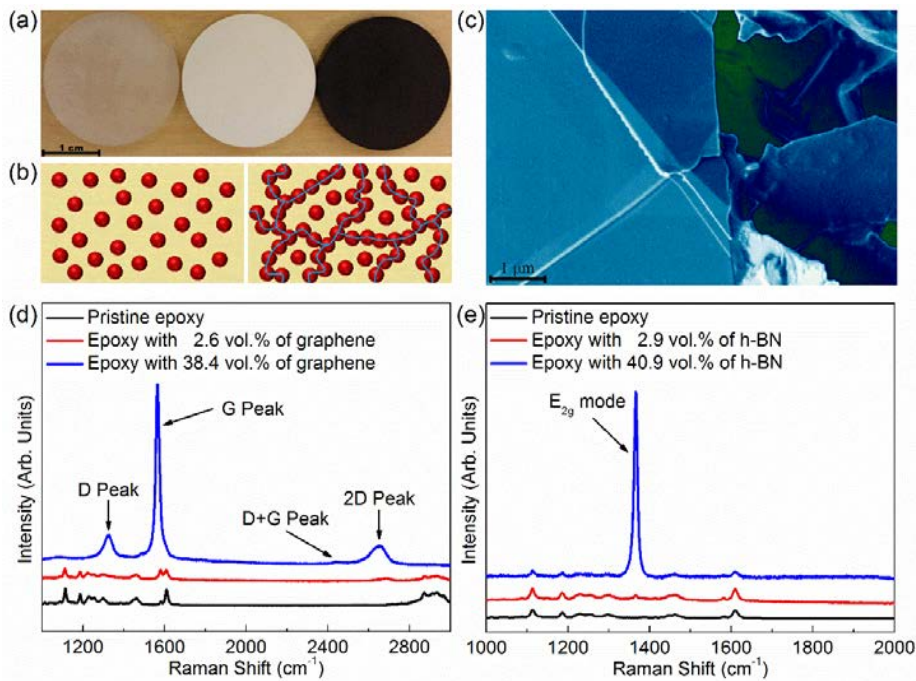

**Figure 1:** (a) Optical image of the pristine epoxy, and epoxy with the loading of 18 vol. % and 19 vol. % of graphene and *h*-BN fillers, respectively. Note a distinctively black color of graphene composite as opposed to the white color of *h*-BN composite. (b) Schematic of the composite with the low (left) and high (right) volume fraction of fillers. (c) Scanning electron microscopy image of the epoxy composite with 45 vol. % of graphene fillers. The microscopy image of the high-loading composites shows clearly the overlapping of graphene fillers inside the epoxy matrix. The overlapping fillers confirm the formation of the percolation network at this high loading fraction of graphene. Raman spectra of (d) pristine epoxy and epoxy with the low and high graphene filler loading fraction; and (e) pristine epoxy and epoxy with the low and high *h*-BN filler loading fraction. The peak at 1366 cm$^{-1}$ is the E$_{2g}$ vibration mode of *h*-BN. In both types of composites, the higher loading of fillers resulted in the increased intensity of the characteristic graphene or *h*-BN phonon peaks allowing for additional verification of the composition of the samples.

The thermal conductivity, $K$, at room temperature (RT) was measured using the transient "hot disk" method[13,51,52]. Details of the measurements are provided in the Methods section and Supplementary Figure 3. Figure 2 (a-b) shows the thermal conductivity as a function of the filler loading fraction, $f$, for composites with graphene and *h*-BN, respectively. One can see that the thermal conductivity enhancement is stronger in composites with graphene than that with *h*-BN. The maximum thermal conductivity enhancements of ×51 and ×24 are achieved for the epoxy composites with graphene $(f = 43 \, \text{vol.} \%)$ and *h*-BN $(f = 45 \, \text{vol.} \%)$, respectively (Supplementary Figure 4 and Supplementary Table 1). Better performance of graphene as the filler material can be explained by its higher intrinsic thermal conductivity, which substantially exceeds that of $h$-BN[1,3,4,53–62]. The intrinsic thermal conductivity of large graphene layers was reported to be in the range from $2000~\mathrm{Wm^{-1}K^{-1}}$ to $5000~\mathrm{Wm^{-1}K^{-1}}$ near RT[1,3,6]. The experimental values reported for thermal conductivity of few layer $h$-BN were found to vary from $\sim230~\mathrm{Wm^{-1}K^{-1}}$ to $\sim480~\mathrm{Wm^{-1}K^{-1}}$ at RT[56–59]. Theoretical calculations reports the thermal conductivity of single layer to few layer $h$-BN in the range from $\sim400~\mathrm{Wm^{-1}K^{-1}}$ to $\sim1000~\mathrm{Wm^{-1}K^{-1}}$ [Refs. 53–55,60–62]. The overall functional dependence of the thermal conductivity of the composites on the loading fraction is the same for both fillers – electrically conducting graphene and electrically insulating $h$-BN. This fact can be explained by the negligible contribution of electrons to heat conduction of graphene near RT[2].

The thermal conductivity depends approximately linearly on the loading fraction for $f_T \lesssim 30~\mathrm{vol.\,\%}$ in graphene composites and $f_T \lesssim 23~\mathrm{vol.\,\%}$ in $h$-BN composites. Above these loading fractions, the dependence become super-linear (Supplementary Figures 5 and 6), indicating the onset of the thermal percolation transport regime[26,28–31,63–66]. The change in the thermal conductivity trend is well resolved for both types of the fillers. The functional dependence $K(f)$ is very different from that in graphite suspension[29] and in the one available prior report of the thermal percolation in graphene[30]. We did not observe singularities in $K(f)$ or $\delta K(f)/\delta f$ dependences but rather an onset of deviation from linear trend. The measured electrical resistivity in the same composites with graphene revealed the electrical percolation threshold at $f_E \approx 10~\mathrm{vol.\,\%}$ - substantially lower loading than $f_T \approx 30~\mathrm{vol.\,\%}$. Another important observation is that despite the large difference in the intrinsic thermal conductivities of graphene, $K_G$, and that of $h$-BN, $K_{h-BN}$ (the ratio $K_G/K_{h-BN} \geq 5$), the thermal percolation is achieved at approximately the same loading fraction. We explain it by the similar lateral dimensions, thicknesses, geometry and flexibility of the graphene and $h$-BN fillers. Below we discuss the results in more detail within the framework of the Lewis-Nielsen model[49,50].

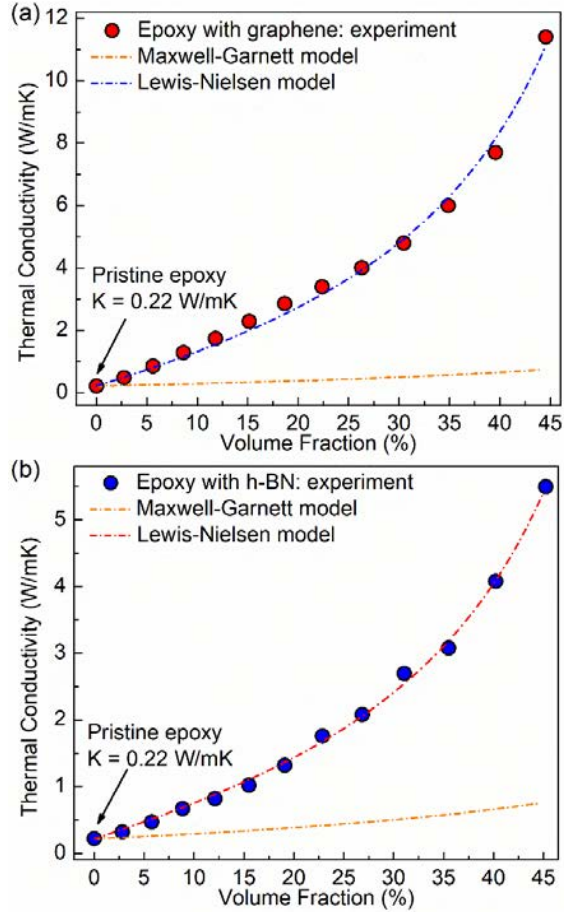

**Figure 2:** Thermal conductivity of the epoxy composites with (a) graphene and (b) *h*-BN fillers over a wide range of the filler loading fraction. The thermal conductivity depends approximately linear on the loading fraction till $f_T \approx 30 \, \text{vol.} \%$ in graphene composites and $f_T \approx 23 \, \text{vol.} \%$ in *h*-BN composites. Above these loading fractions the dependence become super-linear, indicating the onset of the *thermal* percolation transport regime. The maximum thermal conductivity enhancements of ×51 and ×24 are achieved for the epoxy composites with graphene ($f = 43 \, \text{vol.} \%$) and *h*-BN ($f = 45 \, \text{vol.} \%$), respectively. The dash-dot and dash lines correspond to the Lewis-Nielsen and Maxwell-Garnett models, respectively.

The obtained thermal conductivity of epoxy composite with randomly oriented graphene fillers, $K \approx 11 \, \text{Wm}^{-1}\text{K}^{-1}$ at $f = 45 \, \text{vol.} \%$ is rather impressive, and exceeds that of the commercially available TIMs with a similar matrix. This value was obtained with commercial graphene, without additional processing steps. The thermal conductivity of the composites can be further increased at a fixed loading via optimization of the filler lateral sizes and thicknesses[11]. The fillers with the lateral sizes exceeding the phonon mean free path (MFP) in a given material are more efficient in heat conduction (see Supplementary Figure 7 and Supplementary Table 2). However, if the lateral dimensions become too large, the fillers can start folding and rolling, reducing the positive effect. The filler thickness also has an optimum range. Single layer graphene (SLG) has the highest *intrinsic* thermal conductivity[1,3]. However, the thermal conductivity of SLG also suffers the most from the interaction with the matrix material[3]. At the opposite extreme, FLG with the thickness approaching milled graphite, becomes a less efficient filler because it loses its mechanical flexibility. The latter results in weaker coupling to the matrix, *i.e.* larger thermal contact resistance. Achieving the high loading of fillers required the use of commercial graphene and *h*-BN powders with limited control of the thicknesses and lateral dimensions. For this reason, the task of filler size optimization for attaining the maximum thermal conductivity enhancement is reserved for a future study.

We now turn to finding the best theoretical model which can describe the thermal conductivity of the composites over a wide range of $f$. It is needed in order to elucidate the mechanism of heat conduction in the high loading composites above the thermal percolation. Such a model would also be useful for practical applications of composites with graphene or *h*-BN fillers. We have attempted to fit the experimental data with several of the most common models, including the Maxwell-Garnett[67], Lewis-Nielsen[49,50], Agari[68], and others[69]. Previously, it was shown that the Maxwell-Garnett effective medium model (EMM), with the correction for the thermal boundary resistance (TBR) between the fillers and matrix, works well for graphene composites with the low loading $f \leq 10 \, \text{vol.} \%$ [Ref. 11]. We found that the EMM approach does not work well for the graphene or *h*-BN composites over the examined range of the filler loading (see Figure 2 (a-b)). It was also not possible to find proper fitting parameters for the geometrical mean model to match the experimental results (see Supplemental Figures 5 and 6). We succeeded with the semi-empirical model of Lewis-Nielsen, which provided the best fitting to the experimental data over the entire range of the loading fractions (Figure 2 (a-b)). This model explicitly takes into account the effect of the shape, packing of the particles, and, to some degree, their orientation with respect to the heat flux. In the framework of this model, the thermal conductivity is expressed as[50]

$$K/K_m = (1 + ABf)/(1 - B\psi f). \tag{1}$$

Here, the constant $A$ is related to the generalized Einstein coefficient $k_E$ as $A = k_E - 1$, and it depends on the shape of the fillers and their orientation with respect to the heat flow. The parameter $B = (K_f/K_m - 1)/(K_f/K_m + A)$, takes into account the relative thermal conductivity of the two phases: the fillers $(K_f)$ and the base matrix $(K_m)$, respectively. The parameter $\psi = 1 + ((1 - \phi_m)/\phi_m^2)f$ relates to the maximum packing fraction $(\phi_m)$ of the fillers. The values of $A$ and $\phi_m$ are well tabulated for several different two-phase systems and can be found in [Refs. 50,69]. If the shape and packing of the fillers are known, the model predicts the thermal conductivity of the composites without other adjustable parameters.

The values of $A$ and $\phi_m$ are not known for graphene, $h$-BN or other quasi-2D flake-like fillers. Therefore, in this study, we used $A$ and $K_f$ as the fitting parameters to gain insight into the thermal transport in composites with the high loading of quasi-2D fillers. The maximum packing fraction $\phi_m$ was assumed to be 0.52, which corresponds to the three dimensional randomly packed fillers inside the polymer matrix[50]. This is a reasonable assumption since as the filler content increases beyond a certain value – the thermal percolation threshold – the fillers overlap with each other, creating randomly dispersed thermally conductive paths. Fitting our experimental data for the epoxy with graphene and $h$-BN fillers using the Lewis-Nielsen model gave us rather surprising results. First, the *apparent* thermal conductivity values of $K_f \sim 37\,\text{Wm}^{-1}\text{K}^{-1}$ and $K_f \sim 16\,\text{Wm}^{-1}\text{K}^{-1}$ have been extracted for graphene and $h$-BN fillers, respectively. These values are substantially lower than the intrinsic thermal conductivity of graphene and $h$-BN. Second, the values of $A$ obtained from the fitting for two cases of epoxy with graphene and $h$-BN fillers are rather high – ~61 and ~31, respectively. It should be noted that parameter $A$ depends on the shape and the aspect ratio of the filler, and as the aspect ratio increases, the value of $A$ increases as well. For example, for randomly oriented rods with aspect ratios of 2, 4, 6, 10 and 15 the reported values of $A$ are 1.58, 2.08, 2.8, 4.93 and 8.38, respectively[50]. The large $A$ in our case, demonstrates the creation of the rod-like thermal paths with very high aspect ratios. As $A \to \infty$ the Lewis-Nielsen model converges to the ordinary "rule of mixtures", which is $K = (1 - f)K_m + fK_f$ and as $A \to 0$, the model converges to the "inverse rule of mixtures", which is $1/K = (1 - f)/K_m + f/K_f$. The "rule of mixtures" and the "inverse rule of mixtures" provide the upper and lower bounds of the thermal conductivity of composites. Conceptually, the rule of mixtures considers the heat transfer along the two parallel paths of the fillers and the epoxy. For this reason, in our composites with the high loading of graphene and $h$-BN, described by the model with the large $A$ and $K_f/K_m$ ratio, most of the heat is being transferred by the percolated conductive fillers rather than the matrix.

An interesting question is the meaning of the low values of the *apparent* thermal conductivity of the fillers extracted from our experimental data by using the Lewis-Nielsen model. We argue that the low values are mostly defined by the apparent cross-plane thermal conductivity of the quasi-2D fillers rather than the detrimental effect of the matrix and defects, resulting in the decrease of the in-plane thermal conductivity of the fillers. As follows from the discussion above, in our composites with $f \geq f_T$, the dominant, or significant fraction, of the heat propagates via the network of the thermally conductive fillers. In such a network of the overlapping quasi-2D graphene or $h$-BN fillers, the thermal transport is strongly affected by the out-of-plane thermal conductivity of FLG and $h$-BN fillers and TBR at the overlapping regions where the fillers are either directly attached to each other or separated by a thin epoxy layer. The heat has to propagate from one flake to another across their overlapping region (see the inset in Figure 3-a). The cross-plane thermal conductivity of high-quality FLG can be *two orders of magnitude* lower than the in-plane thermal conductivity of FLG fillers – very close to the average apparent values extracted by fitting the Lewis-Nielsen model to our experimental data. The two competing effects – the increase in heat conduction via creation of the thermally conductive filler pathways inside the polymer matrix and thermal resistance associated with the out-of-plane thermal transport at overlapping regions – explains a rather gradual change of the thermal conductivity of the composites at the thermal percolation limit as opposed to a very abrupt change in electrical conductivity at electrical percolation limit. The thermal boundary resistance at the interface between two fillers or at the interface between the filler and epoxy matrix is also a factor, which prevents an abrupt enhancement of the thermal conductivity of the composite as the filler content exceeds the thermal percolation threshold. One can view the extracted thermal conductivity of the fillers as an average apparent quantity, which has contribution from the cross-plane thermal conductivity, implicitly includes TBR between the two overlapping flakes, and is affected by the filler exposure to the matrix material.

To further confirm the conclusion based on the physical model fitting to the experimental data, we conducted a computational study, solving the steady-state heat diffusion equation for a composite region with two overlapping fillers. The schematic of the system is shown in the inset of Figure 3 (a). We consider two identical graphene fillers with the lateral dimensions of $100 \text{ nm} \times 100 \text{ nm}$ and the thickness of $10 \text{ nm}$, embedded in the epoxy. One 2D filler is on top of the other with the overlapping region of 20 % of the filler's area. The heat flux is calculated as a function of the distance separating two fillers, starting from zero, $i.e.$ the fillers are attached to each other, and going to the maximum distance of $d \approx 2 \text{ nm}$. We consider several cases of the anisotropic thermal conductivity of the fillers defined by the ratio of the in-plane – to –cross-plane thermal conductivity $K_{\parallel-f}/K_{\perp-f} = \beta$. Since many of the fillers are FLG flakes, we use, for simplicity, the bulk graphite thermal conductivity for $K_{\parallel-f} = 2000 \text{ Wm}^{-1}\text{K}^{-1}$. The value of $K_{\perp-f}$ varies to simulate the different degree of anisotropy. The high quality graphite has $K_{\perp-f} = 20 \text{ Wm}^{-1}\text{K}^{-1}$ at RT. One should understand that $K_{\perp-f}$ value can implicitly include TBR at the interface between two fillers or filler-matrix material. It is modeled by taking $K_{\perp-f} = 0.2 \text{ Wm}^{-1}\text{K}^{-1}$. The thermal conductivity of the epoxy matrix is taken as $K_m = 0.2 \text{ Wm}^{-1}\text{K}^{-1}$. The details of the simulations can be found in the Methods.

Figure 3 (a) shows the rate of heat flow, $q$, from the hot to the cold filler as a function of the inter-planar distance, $d$, for different values of the cross-plane thermal conductivity of graphene fillers as shown in the legend. The in-plane thermal conductivity is fixed at $2000 \text{ Wm}^{-1}\text{K}^{-1}$. When the fillers are touching ($d = 0 \text{ nm}$), a lower cross-plane thermal conductivity results in a significant decrease in the heat flow from one filler to another. The decrease in the heat flow confirms the importance of the fillers' cross-plane thermal conductivity on heat transfer in the high loading composites, above the thermal percolation threshold. As the separation distance is increased to 2 nm, the effect of the cross-plane thermal conductivity becomes negligible. This situation is similar to that in dilute samples where there are no connections between the fillers. The observed trend explains why the thermal conductivity depends linearly on the loading at lower concentrations ($f < f_T$) and super-linearly on the loading at higher concentrations ($f > f_T$). At $f \leq f_T$, the heat is carried mostly via the epoxy matrix, in which the thermal conductivity is enhanced by separate islands of the fillers. However, at $f \geq f_T$, the heat is conducted mostly via the connected network of fillers with the apparent thermal conductivity limited by the cross-plane thermal conductivity, which is still an order of magnitude higher than that of the epoxy.

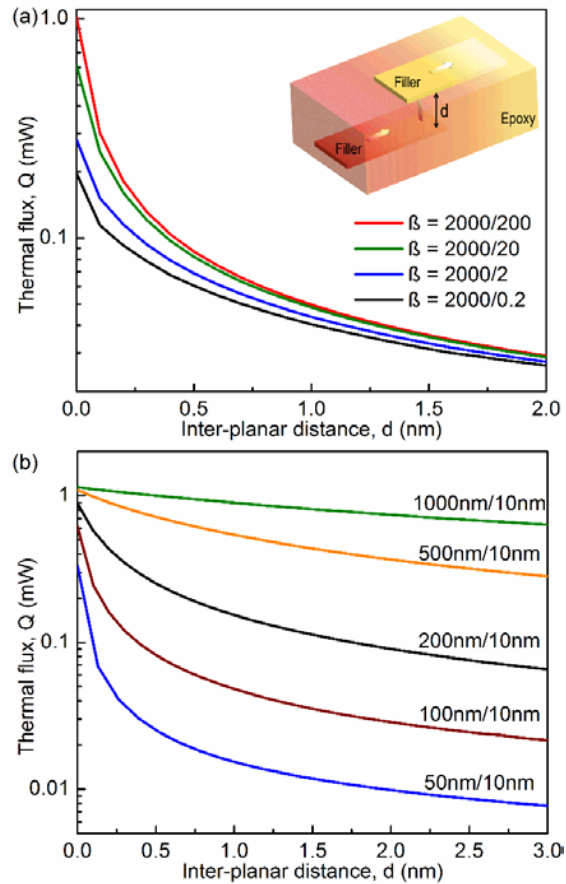

**Figure 3:** (a) Effect of filler's thermal conductivity anisotropy on heat flux across the overlapping fillers as a function of the inter-planar distance between the fillers for 20% overlap at constant filler size aspect ratio AR = 100 nm /10 nm (length/thickness) and filler volume fraction of $f = 28$ vol. %. The cross-plane thermal conductivity of the graphene fillers strongly affects the heat transfer in case of the thermally percolated fillers. The inset shows schematically the heat propagation from one filler to the other (b) The effect of the aspect ratio on heat propagation in the percolated network for 20% overlap between the filler flakes. The results have been obtained for filler loading fraction of $f = 28$ vol. % and thermal conductivity anisotropy of $\beta = 2000/20 = 100$  The increase of the filler lateral size, at fixed thickness,

results in better heat conduction from one filler to another as compared to the case where the fillers are small.

Figure 3 (b) shows the thermal flux for a constant anisotropy of the thermal conductivity of $\beta = 2000/20 = 100$, as a function of the filler's aspect ratio (AR), defined as the ratio of the flake length to its thickness. We assumed the flakes have the same length and width. The increase of the filler lateral size, at fixed thickness, results in better heat conduction from one filler to another as compared to the case where the fillers are small. At very small aspect ratios, *e.g.* AR=5, the heat transfer decreases abruptly as the separation distance $d$ increases to $\sim 1 - 2$ nm. This indicates that the heat dissipates to the epoxy environment, which has very low thermal conductivity. The considered AR values are close to the experimental range, *e.g.* the flakes with micrometer lateral dimensions and 10 nm thickness have AR=100. Practically, the upper bound for the filler size will be defined by the material processing technique and dimensions at which the fillers start to bend and roll over. One should also remember that if the lateral dimension of the filler becomes smaller than the phonon mean-free-path (MFP), its thermal conductivity starts to decrease due to onset of the ballistic phonon transport regime[1–3].

While the percolation threshold loading fraction, $f_T$, is rather obvious in our case as the point where the linear dependence becomes super-linear (see Supplemental Figs 5 and 6), there are no distinctive bends in the functional dependence (*e.g.* the derivative is smooth and continuous). In order to conclusively prove the change in the nature of thermal transport after reaching the percolation regime, we investigated the thermal diffusivity as a function of temperature for the low-loading and high-loading composites. Figure 4 (a-b) shows the thermal diffusivity, $\alpha$, for composites with graphene and *h*-BN fillers. The thermal diffusivity of the samples has been measured using laser flash technique (see Methods section and Supplementary Figure 8). For the pristine epoxy and composites with the low filler content, the thermal diffusivity does not change with increasing temperature. The slope of the curves for the epoxy and epoxy with $\sim 2$ vol. % of graphene and *h*-BN is in the order of $10^{-4}$ mm$^2$s$^{-1}$K$^{-1}$. In the high loading composites, the thermal diffusivity decreases with increasing temperature. In samples with $f \geq 19$ vol. %, the slope of the curves, which characterizes the rate of change of the diffusivity, is $10^{-2}$ mm$^2$s$^{-1}$K$^{-1}$ – two orders of magnitude larger than for the samples with $f \leq 3$ vol. % of the filler loading. The change in the temperature dependence of the thermal diffusivity can be explained from the following considerations. Below the percolation limit, the heat is conducted mostly through the disordered matrix, with only small fraction via the thermally conductive fillers. For this reason, the thermal diffusivity does not depend on temperature, or depends very weakly, as typical for amorphous materials in the examined temperature range. Above the percolation limit, heat mostly or partially travels via the percolation network, made from crystalline fillers such as graphene or $h$-BN. In this case, the temperature, $T$, dependence starts to evolve closer to $1/T$ law, characteristic for crystalline solids above RT. In this sense, the change in $\alpha(T)$ functional dependence can be used as an additional criterion for distinguishing the onset of the percolation transport regime in the high loading composites.

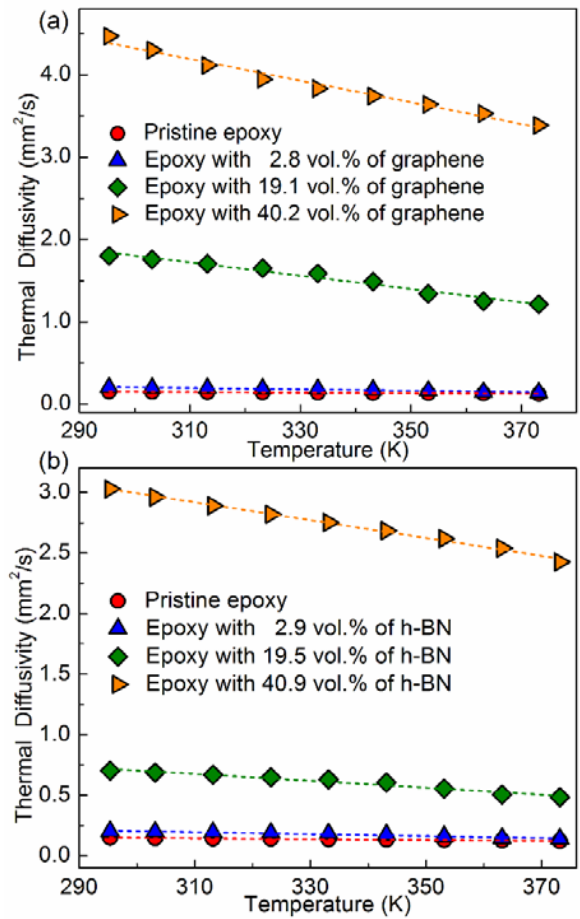

**Figure 4:** Thermal diffusivity of the epoxy composite with (a) graphene and (b) $h$-BN fillers as a function of temperature. For pristine epoxy and the low-filler-content composites, the thermal

diffusely does not change with increasing temperature. The thermal diffusivity reveals a decreasing trend with increasing temperature in the high-filler-content composites with the loading fraction $f \geq f_T$.

The change in the nature of heat conduction above the thermal percolation limit becomes even more clear if one examines the temperature dependence of the thermal conductivity in the composites with the gradually changing filler content, from $f = 0$ to $f > f_T$ (see Figure 5). The thermal conductivity of the pristine epoxy increases slowly and monotonically in the temperature range from $\sim 20\,°\text{C}$ to $\sim 110\,°\text{C}$ as expected for amorphous or disordered electrically insulating materials. As a small loading fraction of graphene (2.7 vol.%) is added to the epoxy, the thermal conductivity shows a maximum at $T_{max} = 100\,°\text{C}$, and then starts to decrease. Addition of more graphene results in the increase of the thermal conductivity and the shift of $T_{max}$ to lower temperatures owing to the gradual change of the structure of the composite from more amorphous to more crystalline due to the filler introduction. The maximum in thermal conductivity functional dependence on temperature, $K(T)$, is reminiscent of that in crystalline solids, although the maximum in our thermally percolated composites happens at substantially higher temperature.

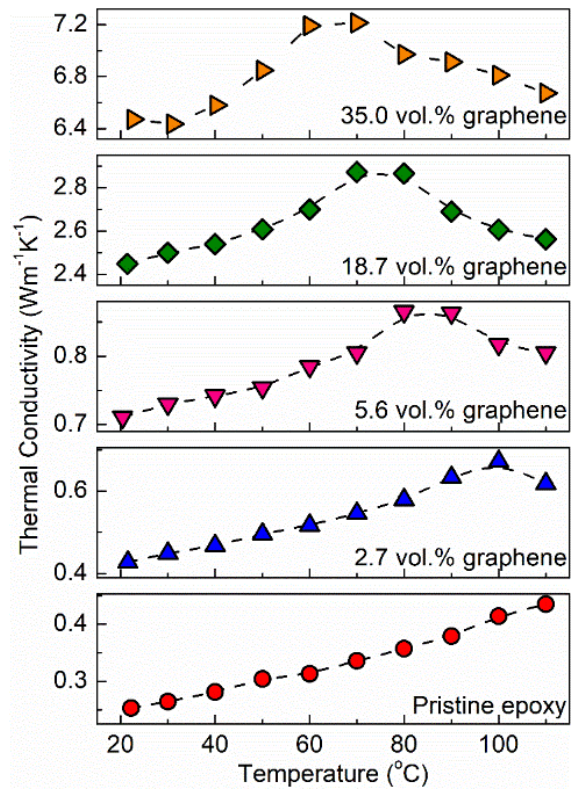

**Figure 5:** Thermal conductivity of the epoxy with graphene fillers as a function of temperature. The thermal conductivity of the pristine epoxy increases gradually with temperature as expected for amorphous or disordered materials. Addition of graphene results in the increase in the thermal conductivity and change in the thermal conductivity's dependence on temperature. Note an appearance of a pronounced maximum in the thermal conductivity and its shift to low temperatures as graphene filler content increases.

In conclusions, we investigated the thermal conductivity and thermal diffusivity of epoxy composites with the high loading fraction of graphene and $h$-BN fillers. It was found that both types of the composites revealed a *distinctive* thermal percolation threshold at a rather high loading fraction $f_T > 20$ vol. %. The changes in thermal transport at high loading fractions were confirmed by the temperature dependence of the thermal diffusivity. The graphene fillers outperformed boron nitride fillers in terms of thermal conductivity enhancement. It was also established that the thermal conductivity of both types of the composites can be best described by the Lewis-Nielsen model. The surprising finding is that in the high loading composites with quasi-2D fillers, the apparent cross-plane thermal conductivity of the fillers can be the limiting factor for heat conduction. The obtained results are important for developing the next generation of the thermal interface materials.

**METHODS**

**Sample Preparation:** The composite samples were prepared by mixing commercially available few-layer graphene (Graphene Supermarket), $h$−BN (US Research Nanomaterials, Inc.), and epoxy (Allied High Tech Products, Inc.). Graphene and $h$−BN flakes were added to the epoxy resin at different mass ratios to prepare composites with varying filler contents. For samples containing a low filler loading fraction, the epoxy resin and the fillers were mixed using the high-shear speed mixer (Flacktek Inc.) at 800 rpm and then 2000 rpm each for 5 minutes. Then, the homogeneous mixture of epoxy and fillers were put inside a vacuum chamber for ~10 minutes in order to extract the trapped air bubbles. The curing agent (Allied High Tech Products, Inc.) was then added in the prescribed mass ratio of 1:12 with respect to the epoxy resin. The solution was twice more mixed and vacuumed following the procedure outlined previously. Finally, the mixture was placed in an oven for ~2 hours at 70º C to cure and solidify.

For the high-volume fraction samples, the graphene and $h$-BN fillers were added to the epoxy resin at several steps. In the first step, $1/3$ of the total loading weight of the filler was added to half of the total weight of the required epoxy resin and was dispersed using a speed mixer at 2000 rpm for 5 minutes. Then, another $1/3$ of the filler was mixed with the rest of the required epoxy resin. The solution was mixed one more time at 2000 rpm but for 10 minutes. The viscous mixture was then degassed in the vacuum chamber for 5 minutes. Afterwards, the rest of the required graphene or $h$-BN was added with the curing agent to the mixture and stirred slowly using a home-made stirrer with the sharp needles. The needles helped to prevent filler agglomeration. At the next step, the composite was mixed at the high rotation speeds of 3500 rpm and 2000 rpm for 15 seconds and 10 minutes, respectively. The homogenous mixture was gently pressed and left in the oven at 70 ℃ for 2 hours to cure. Using this procedure, several samples with a high filler loading of up to $f \approx 45$ vol. % were successfully prepared.

**Mass Density Measurements**: To compare our thermal conductivity experimental results with available theoretical models, we converted the filler mass fraction ratio ($\varphi$) to the volume fraction ($f$) according to $f = \varphi\rho_m/(\varphi\rho_m + (1-\varphi)\rho_f)$ equation. Here, $\rho_m$ and $\rho_f$ are the density of the epoxy and the filler, respectively. However, the density of the graphene and $h$-BN fillers can vary depending on the production method and possible impurities[3,70]. For this reason, we measured accurately the mass ($m$) and volume ($V$) of the disk-shaped samples and calculated the density of the composite samples according to $\rho = m/V$. Following an iterative procedure, we first assumed that the density of the epoxy, graphene and $h$-BN are $1.16, 2.26$ and $2.16$ gcm$^{-3}$. We converted the filler mass fraction of the samples to the filler volume fraction according to the aforementioned equation. We plotted the density of the composites versus the obtained volume fraction data and fitted the experimental data using a linear regression (see Supplementary Figures 1 and 2). Based on the rule of mixtures for composites, $\rho = f(\rho_f - \rho_m) + \rho_m$, the $y-$intercept of the fitted line and the slope will be equal to the density of the epoxy ($\rho_m$) and the difference between the filler and epoxy densities ($\rho_f - \rho_m$), respectively. Calculating the volume fraction based on the new values and following an iterative procedure discussed above, we extracted the exact values of the density for the fillers and the epoxy matrix. In our case, for epoxy with graphene fillers the density of the graphene and epoxy was calculated as 1.16 gcm$^{-3}$ and 2.16 gcm$^{-3}$, respectively. For the epoxy with $h$-BN fillers, the density of the $h$-BN and epoxy was calculated as 1.17 gcm$^{-3}$ and 2.07 gcm$^{-3}$, respectively.

**Thermal Conductivity Measurements:** The thermal conductivity was measured using the transient plane source (TPS) "hot disk" technique. In TPS method, an electrically insulated flat nickel sensor is placed between two pieces of the substrate. The sensor is working as the heater and thermometer simultaneously. A current pulse is passed through the sensor during the measurement to generate the heat wave. Thermal properties of the material are determined by recording temperature rise as a function of time. The time and the input power are chosen so that the heat flow is within the sample boundaries and the temperature rise of the sensor is not influenced by the outer boundaries of the sample. More details on the measurement procedures can be found in the Supplemental Materials and our prior reports on other material systems[13,51,52,71–80].

**Thermal Diffusivity Measurements:** The measurements of the thermal diffusivity were performed by the transient "laser flash" (LFA) technique (NETZSCH LFA 447). The LFA technique uses a flash lamp, which heats the sample from one end by producing shot energy pulses. The temperature rise is determined at the back end with the infrared detector. The output of the temperature detector is amplified and adjusted for the initial ambient conditions. The recorded temperature rise curve is the change in the sample temperature resulting from the firing of the flash lamp. The magnitude of the temperature rise and the amount of the light energy are not required for the diffusivity measurement. Only the shape of the transient curve is used in the analysis. More details on the measurement procedures can be found in the Supplemental Materials and our prior reports on other material systems[9,11,80,81].

**Numerical Simulations**: The heat conduction in the system consisting of the epoxy matrix and graphene fillers was calculated using the finite element method (FEM) as implemented in COMSOL Multiphysics package. The few-layer graphene fillers were considered to be thermally anisotropic planes with in-plane (out-of-plane) thermal conductivity $K_{\parallel}$ ($K_{\perp}$). The epoxy matrix was assumed to have uniform thermal conductivity $K_m$ of 0.2 Wm$^{-1}$K$^{-1}$. The inter-planar distance, $d$, between the fillers were varied from zero to 10 nm. The heat transfer in the regions of the matrix (m) and graphene fillers (g) was described by their respective thermal conductivities $K_{m,g}$, as $\nabla \cdot \left( K_{m,g} \nabla T_{m,g} \right) = 0$. The fixed temperature boundary conditions were applied along the left and right faces of the simulation domain while all other faces were assumed to be adiabatic, $\partial T_m / \partial n_m = 0$ where $n_m$ is the outward normal to the surface.

## *Acknowledgements*

This work was supported, in part, by the National Science Foundation (NSF) through the Emerging Frontiers of Research Initiative (EFRI) 2-DARE award 1433395: Novel Switching Phenomena in Atomic $MX_2$ Heterostructures for Multifunctional Applications, and by the UC – National Laboratory Collaborative Research and Training Program. A.A.B. also acknowledges NSF award 1404967: CDS&E Collaborative Research: Genetic Algorithm Driven Hybrid Computational – Experimental Engineering of Defects in Designer Materials.

## Contributions

A.A.B. and F.K. conceived the idea of the study. A.A.B. coordinated the project and contributed to the experimental and theoretical data analysis; F.K. conducted data analysis and assisted with the thermal measurements; Z.B. prepared the composites and performed thermal and electrical conductivity measurements; J.S.L. and R.S. conducted materials characterization, thermal conductivity and thermal diffusivity measurements; B.D. performed numerical modeling. E.A. carried out Raman spectroscopy and related materials characterization. R.L. contributed to the theoretical and computational data analysis. A.A.B. led the manuscript preparation. All authors contributed to writing and editing of the manuscript.